# Review of flexible and transparent thin-film transistors based on zinc oxide and related materials[*]

Zhang Yong-Hui(张永晖)[1], Mei Zeng-Xia(梅增霞)[1†], Liang Hui-Li(梁会力)[1], Du Xiao-Long(杜小龙)[1, 2†]

[1]*Key Laboratory for Renewable Energy,Beijing National Laboratory for Condensed Matter Physics,Institute of Physics,*
*Chinese Academy of Sciences, Beijing 100190, China*
[2]*School of Physical Sciences, University of Chinese Academy of Sciences, Beijing 100190, China*

Flexible and transparent electronics presents a new era of electronic technologies. Ubiquitous applications involve wearable electronics, biosensors, flexible transparent displays, radio-frequency identifications (RFIDs), etc.Zinc oxide (ZnO) and related materials are the most commonly used inorganic semiconductors in flexible and transparent devices, owing to their high electrical performance, together with low processing temperature and good optical transparency.In this paper, we review recent advances in flexible and transparent thin-film transistors (TFTs) based on ZnO and related materials.After a brief introduction, the main progresses on the preparationof each component (substrate, electrodes, channel and dielectrics) are summarized and discussed. Then, the effect of mechanical bending on electrical performance was highlighted. Finally, we suggest the challenges and opportunities infuture investigations.

## 1. Introduction

The last thirteen years have witnessed the rise of flexible and transparent electronics.Since Hoffman et al. demonstrated the first fully transparent zinc oxide thin-film transistor (ZnO TFT) in 2003,[1]numerous important works have been reported.[2–21] The typical applications involve active-matrix flexible or transparent displays, logic circuits, electronic skins, bio-sensorsandwearable devices.Owing to their mechanical flexibility, optical transparency, light weight, low production cost, low power consumption and, above all, high electrical performance, these devices have drawn broad interests in both academy and industrial circles. A wide range of diverse applications of flexible and transparent TFTs based on ZnO related materials

[*]Project supported by theNationalScienceFoundation of China(GrantsNos.61306011, 11274366, 51272280,11674405, and11675280)
[†]Corresponding authors. E-mails: zxmei@iphy.ac.cn and xldu@iphy.ac.cn

are listed in Fig. 1.

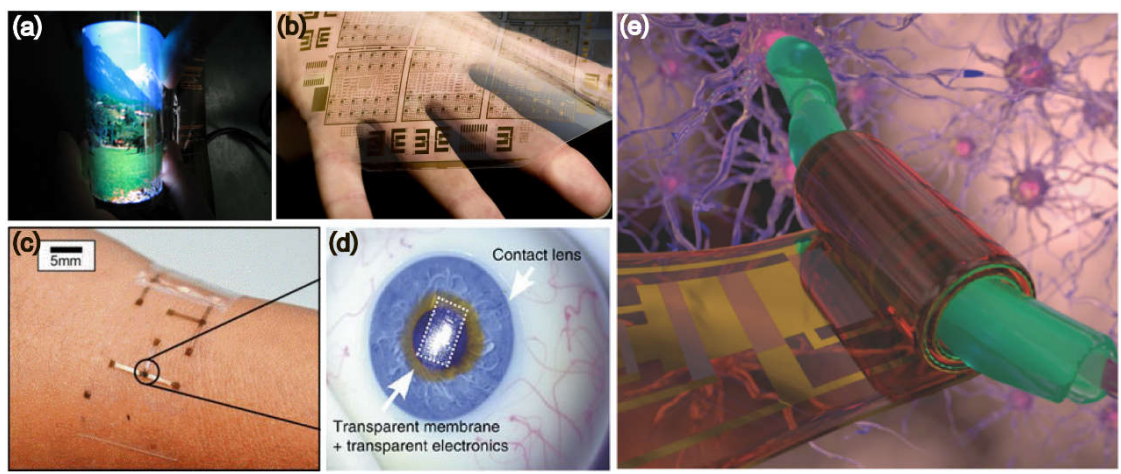

Fig.1 Versatile applications of ZnO related TFTs. (a)6.5 in. flexible full-color display driven by indium gallium zinc oxide (IGZO) TFTs.[6](b) Flexible transparent IGZO circuits.[7](c)Electronic skin based on IGZO TFT.[22](d)Smart contact lens based on IGZO TFTs.[23] (e) Biomimetic neuronal microelectronics based on IGZO TFT.[16]

Organic and hydrogenated amorphous silicon (a-Si:H) TFTs have also been demonstrated but their applications were limited by the low mobility of the conductive channels. ZnO and related materials have thus emerged as promising candidates for channel materials in flexible and transparent TFTssince the birth of this subject field at 2003-2004.[1,4]Comparedwith other inorganic wide bandgap semiconductors, such as gallium nitride (GaN) and silicon carbide (SiC), the great advantage of ZnO materials for flexible devices is their low synthesis temperature, which is exactly the most importantrequirementin flexible device fabrication processes.[24](For details, see section 2.)The biocompatibility of ZnO materials also makes them perfectly desirable for medicaland health-care applications as shown in Fig. 1c-e.In addition, the feasibility of modulatingthe electrical properties via doping or alloying with other elements offers the opportunities to adjust device performancesandthus own diverse functionalities.The most commonly used alloys are indium zinc oxide (IZO),IGZO, zinc tin oxide (ZTO), zinc indium tin oxide (ZITO), and magnesium zinc oxide (MZO). Table 1 lists some of the state of the art flexible transparent TFTs based on versatile ZnO and related materials in the past 6 years.

Table 1 Some of the state of the art flexible transparent TFTs based on ZnO and related materials from 2010-2016. ("--" means not mentioned or not clear in the literature.)

| Material | Technique | $T_{dep}/T_{post}$ (°C) | Substrate | Dielectric | $\mu$ (cm²V⁻¹s⁻¹) | On/off | Reference | Year |
|---|---|---|---|---|---|---|---|---|
| ZnO | Spin-coating | --/135 | PEN | RSiO$_{1.5}$ | 0.07 | 10⁴ | Fleischhaker et al.[25] | 2010 |
| ZnO | ALD | 150/-- | PI | Al$_2$O$_3$ | 3.07 | 10² | Cherenack et al.[26] | 2010 |
| ZnO | Sputtering | RT/350 | PDMS | SiO$_2$ | 1.3 | 10⁶ | Park et al.[27] | 2010 |
| ZnO | Hydrothermal | 90/100 | PET | PMMA | 7.53 | 10⁴ | Lee et al.[28] | 2010 |
| ZnO | Spin-coating | RT/200 | PI | SiO$_2$ | 0.35 | 10⁶ | Song et al.[29] | 2010 |
| ZnO | PEALD | 200/-- | PI | Al$_2$O$_3$ | 20 | 10⁷ | Zhao et al.[30] | 2010 |
| IGZO | Sputtering | RT/-- | PI | Al$_2$O$_3$ | 12.85 | 10⁶ | Cherenack et al.[26] | 2010 |
| IGZO | Sputtering | --/-- | PET | BST/PMMA | 10.2 | 10⁶ | Kim et al.[31] | 2010 |
| IGZO | PLD | RT/-- | PET | Parylene | 3.2 | 10⁷ | Nomura et al.[32] | 2010 |
| IGZO | Sputtering | RT/200 | PI | HfLaO | 22.1 | 10⁵ | Su et al.[33] | 2010 |
| ZITO | PLD | --/-- | PET | Ta$_2$O$_5$/SiO$_x$ | 20 | 10⁵ | Liu et al.[34] | 2010 |
| ZITO | PLD | RT/-- | PET | v-SAND | 110 | 10⁴ | Liu et al.[35] | 2010 |
| ZITO | Sputtering | RT/-- | Polyarylate | Al$_2$O$_3$ | 16.93 | 10⁹ | Cheong et al.[36] | 2010 |
| ZnO | Spin-coating | RT/MW140 | PES | SiO$_2$ | 0.57 | 10³ | Jun et al.[37] | 2011 |
| ZnO | Spin-coating | --/150 | PES | Hybrid | 0.142 | 10⁴ | Jung et al.[38] | 2011 |
| IGZO | Sputtering | 200/220 | PI | SiO$_2$ | 19.6 | 10⁹ | Mativenga et al.[39] | 2011 |
| IGZO | Sputtering | --/150 | PEN | Al$_2$O$_3$ | 17 | 10⁸ | Tripathi et al.[7] | 2011 |
| IGZO | Sputtering | 200/220 | PI | SiO$_2$ | 19 | 10⁹ | Mativenga et al.[39] | 2011 |
| IGZO | -- | --/-- | PI | Al$_2$O$_3$ | 13.7 | 10⁷ | Kinkeldei et al.[40] | 2011 |
| IGZO/IZO | Sputtering | 40/200 | PEN | SiO$_2$ | 18 | 10⁹ | Mars et al.[41] | 2011 |
| IGZO | Sputtering | RT/-- | PI | Al$_2$O$_3$ | 14.5 | 10⁷ | Münzenrieder et al.[42] | 2012 |
| IGZO | Sputtering | RT/-- | PEN | PVP | 0.43 | 10⁵ | Lai et al.[43] | 2012 |
| IGZO | Sputtering | RT/-- | PU | Al$_2$O$_3$ | 9.36 | 10⁵ | Erbet al.[44] | 2012 |
| IGZO | Sputtering | RT/-- | PI | PVP | 3.6 | 10⁴ | Kim et al.[45] | 2012 |
| IGZO | Sputtering | --/PN254nm | PAR | Al$_2$O$_3$ | 7 | 10⁸ | Kim et al.[46] | 2012 |
| ZnO | Sputtering | 100/-- | PET | HfO$_2$ | 7.95 | 10⁸ | Ji et al.[47] | 2013 |
| ZnO | Spin-coating | --/200 | PET | c-PVP | 0.09 | 10⁵ | Kim et al.[48] | 2013 |
| ZnO | Printing | --/250 | PI | Ion-gel | 1.67 | 10⁵ | Hong et al.[49] | 2013 |
| ZnO | Spin-coating | --/160 | PEN | Al$_2$O$_3$-ZrO$_2$ | 5 | 10⁴ | Lin et al.[50] | 2013 |
| IZO | Spin-coating | RT/280 | PI | Zr-Al$_2$O$_3$ | 51 | 10⁴ | Yang et al.[51] | 2013 |
| IZO | Sputtering | RT/RT | PET | SiO$_2$ | 65.8 | 10⁶ | Zhou et al.[52] | 2013 |
| IZO:F | Spin-coating | RT/200 | PEN | Al$_2$O$_3$ | 4.1 | 10⁸ | Seo et al.[53] | 2013 |
| IGZO | Sputtering | RT/110 | PET | c-PVP | 10.2 | 10⁶ | Hyung et al.[54] | 2013 |
| IGZO | Sputtering | --/-- | PC | Y$_2$O$_3$/TiO$_2$/Y$_2$O$_3$ | 32.7 | 10⁶ | Hsu et al.[55] | 2013 |
| IGZO | Sputtering | --/-- | PC | GeO$_2$/TiO$_2$/GeO$_2$ | 8 | 10⁷ | Hsu et al.[56] | 2013 |
| IGZO | Sputtering | RT/RT | PC | SiO$_2$/TiO$_2$/SiO$_2$ | 76 | 10⁵ | Hsu et al.[57] | 2013 |
| IGZO | Sputtering | RT/-- | PI | Al$_2$O$_3$ | 7.3 | 10⁹ | Münzenrieder et al.[58] | 2013 |
| IGZO | Sputtering | RT/-- | PI | Al$_2$O$_3$ | 8.3 | 10⁸ | Münzenrieder et al.[59] | 2013 |
| IGZO | Sputtering | RT/-- | PI | Al$_2$O$_3$ | 18.3 | 10⁸ | Zysset et al.[60] | 2013 |
| IGZO | Printing | --/-- | PDMS | SiO$_2$ | 15 | 10⁶ | Sharma et al.[61] | 2013 |
| GNS/IGZO | Spin-coating | --/500 | Thin glass | Ta$_2$O$_5$ | 23.8 | 10⁶ | Dai et al.[62] | 2013 |

| ZnO | Sputtering | --/MW | PES | $Al_2O_3$ | 1.5 | $10^6$ | Park et al.[63] | 2014 |
|---|---|---|---|---|---|---|---|---|
| IZO | Spin-coating | RT/350 | PI | K-PIB | 4.1 | $10^5$ | Wee et al.[64] | 2014 |
| IGZO | Sputtering | --/-- | PI | $Al_2O_3$/ $SiO_2$ | 9 | $10^7$ | Chen et al.[65] | 2014 |
| IGZO | Sputtering | --/160 | PEN | $Al_2O_3$ | 11.2 | $10^9$ | Xu et al.[66] | 2014 |
| IGZO | Sputtering | 150/150 | PEN | $Al_2O_3$ | 12.87 | $10^9$ | Xu et al.[67] | 2014 |
| IGZO | Sputtering | RT/-- | PEN | Nanocomposite | 5.13 | $10^5$ | Lai et al.[68] | 2014 |
| IGZO | Sputtering | RT/250 | PI | $Al_2O_3$ | 14.88 | $10^8$ | OK et al.[69] | 2014 |
| IGZO | Sputtering | RT/190 | PEN | $SiO_2$ | 8 | $10^7$ | Nakajima et al.[70] | 2014 |
| IGZO | Spin-coating | RT/350 | PI | $Al_2O_3$ | 84.4 | $10^5$ | Rim et al.[71] | 2014 |
| IGZO | Sputtering | --/-- | Parylene | $Al_2O_3$ | 11 | $10^4$ | Salvatore et al.[23] | 2014 |
| IGZO | Sputtering | RT/-- | PI | $Al_2O_3$ | 7.5 | $10^7$ | Petti et al.[72] | 2014 |
| IGZO | Sputtering | RT/-- | PI | $Al_2O_3$ | 10.5 | $10^8$ | Münzenrieder et al.[73] | 2014 |
| IGZO | Sputtering | --/300 | Thin glass | $Si_3N_4$ | 9.1 | $10^8$ | Lee et al.[74] | 2014 |
| ZnO | PEALD | 200/-- | PI | $Al_2O_3$ | 12 | $10^8$ | Li et al.[75] | 2015 |
| IZO | Sputtering | --/-- | PET | $SiO_2$ | 12 | $10^5$ | Liu et al.[76] | 2015 |
| IGZO | Sputtering | --/-- | -- | $Si_3N_4$ | 6.5 | $10^6$ | Kim et al.[77] | 2015 |
| IGZO | Sputtering | RT/-- | PI | $Al_2O_3$/ $SiO_2$ | 4.93 | 5 | Honda et al.[78] | 2015 |
| IGZO | Spin-coating | --/PN254nm | PI | ZAO | 11 | $10^9$ | Jo et al.[79] | 2015 |
| IGZO | Sputtering | --/160 | PEN | $SiO_2$ | 7 | $10^7$ | Motomura et al.[80] | 2015 |
| IGZO | Sputtering | RT/200 | PDMS | P(VDF-TrFE) | 21 | $10^7$ | Jung et al.[81] | 2015 |
| IGZO | Sputtering | RT/-- | PI | $Al_2O_3$ | 17 | $10^5$ | Karnaushenko et al.[16] | 2015 |
| IGZO | Sputtering | RT/150 | PVA | $SiO_2$/$Si_3N_4$ | 10 | $10^6$ | Jin et al.[82] | 2015 |
| IGZO | Sputtering | RT/180 | PEN | $Al_2O_3$ | 15.5 | $10^9$ | Park et al.[83] | 2015 |
| IGZO | Sputtering | --/-- | PI | $Al_2O_3$ | 0.2 | $10^4$ | Petti et al.[84] | 2015 |
| IGZO | Sputtering | --/180 | PEN | $Si_3N_4$ | 13 | $10^8$ | Tripathiet al.[85] | 2015 |
| IGZO/TO | Sputtering | --/-- | PC | $TiO_2$/$HfO_2$ | 61 | $10^5$ | Hsu et al.[86] | 2015 |
| ZnO | Sputtering | --/225 | PI | $HfO_2$ | 1.6 | $10^6$ | Li et al.[87] | 2016 |
| ZnO | Sputtering | RT/RT | PEN | $Al_2O_3$ | 11.56 | $10^8$ | Zhang et al.[88] | 2016 |
| IZO | Sputtering | --/300 | PI | $Al_2O_3$ | 6.64 | $10^7$ | Zhang et al.[89] | 2016 |
| IZO | SCS | 275/-- | Polyester | $Al_2O_3$/ $ZrO_2$ | 3.9/6.2 | $10^4$ | Wang et al.[90] | 2016 |
| IGZO | Sputtering | RT/-- | PI | $SiO_2$ | 12 | $10^7$ | Park et al.[91] | 2016 |
| IGZO | Sputtering | RT/200 | PDMS | P(VDF-TrFE):PMMA | 0.35 | $10^4$ | Jung et al.[92] | 2016 |
| IGZO | Spin-coating | --/ PN254nm | PI | $Al_2O_3$ | 5.41 | $10^8$ | Kim et al.[93] | 2016 |
| IGZO | Sputtering | --/-- | PES | $Al_2O_3$ | 71.8 | $10^8$ | Oh et al.[94] | 2016 |
| ZTO | Inkjet printing | 30/300 | PI | $ZrO_2$ | 0.04 | $10^3$ | Zeumault et al.[95] | 2016 |
| ZITO | Sputtering | 300/200 | PI | $SiO_2$ | 32.9 | $10^9$ | Nakata et al.[96] | 2016 |

Besides TFTs, the operation of a flexible transparent circuits also needs high-performance thin film diodes. However, research on flexible transparent diodes are quite limited despite the great progress made in TFTs as shown in Table 1. The limited reports can be categorized into 4 types: (1) *pn* heterojunction diode.[97–102] As most of the wide-bandgap semiconductors are n-type conductive, a proper p-type wide-bandgap material must be chosen wisely to form a large built-in potential barrier. (2) Schottky junction diode.[103–106] The electron affinities (χ) of most wide-bandgap materials are more than 4 eV, thus only a small Schottky barrier height (*SBH*) could be formed with non-noble metals. (3) Metal-insulator-semiconductor (MIS) diode.[107] As in Schottky diode, a large difference between metal work function ($\Phi_M$) and semiconductor affinity ($\chi$) is necessary to achieve a large rectification ratio. (4) Metal-insulator-metal (MIM) diode.[108,109] Actually, it is not easy for MIM diodes to be applied in transparent circuits because of the difficulties to find two kinds of transparent electrodes with large work function difference. (5) Self-switching diode (SSD).[110] Besides small rectification ratio, this kind of device involves nanofabrication process, which may bring high costs and challenges in technological compatibility. Zhang et al. recently demonstrated high-performance flexible fully transparent ZnO diodes with a high rectification ratio of $10^8$ by using a diode-connected TFT architecture as shown in Fig. 2a.[88] The device fabrication procedure is the same as standard TFTs (Fig. 2b). Both of the devices on polyethylene naphthalate (PEN) and quartz substrates are optically transparent (inset of Fig. 2c), with the whole devices (including the substrates) exhibiting a transmittance over 80% in full visible spectral range (Fig.2c). Most importantly, the devices exhibit a high rectification ratio of $5 \times 10^8$ (Fig.2d) under flat and bent states (inset of Fig. 2d), which is 3-4 orders larger than conventional junction diodes. This work has broadened the application value of flexible transparent TFTs and provided a solution to flexible fully transparent diodes which may be used for reference in flexible transparent TFTs based on other materials.

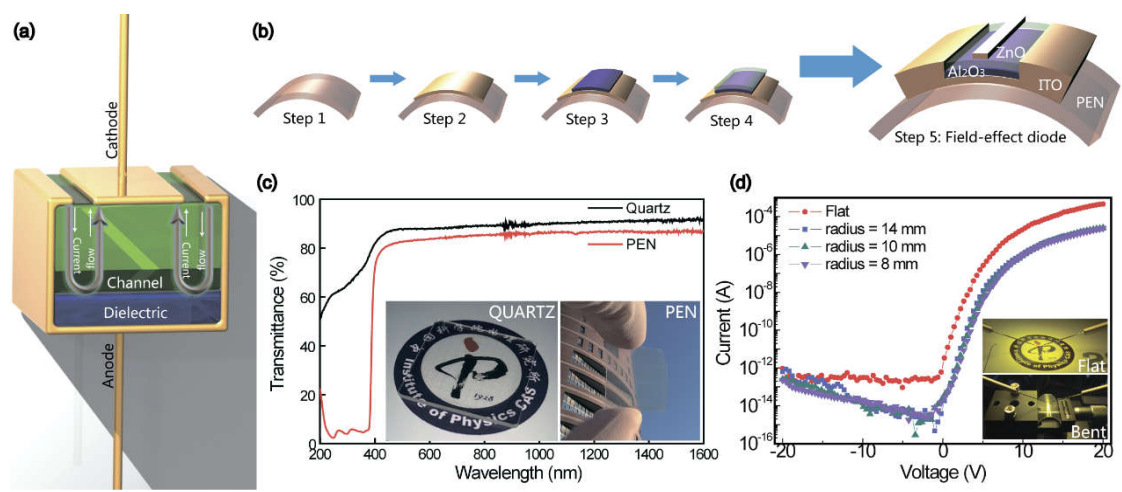

Fig.2 Device structure and performance of flexible transparent ZnO diode. (a) Conceptualized structure diagram, which features two-terminal configuration. (b) Fabrication procedure of field-effect diode. (c) Optical transmittance spectra of devices on glass and PEN substrates with transmittance over 80% in visible spectral range. The insets show the photographs of these two devices. (d) Current-voltage (I-V) characteristics of field-effect diode with a high rectification ratio of $5 \times 10^8$ while flat and bent. Insets are photographs of device under test.[88]

Thanks to the rapid development of science and technology in flexible transparent electronics, many wonderful products are very close to achieving commercial-productions.[111,112] In fact, IGZO panels have already been used in iPad

Airand iPad Pro products, and Apple is considering IGZO panel for its new iPhone in late 2017.As for flexible transparent display and other more applications, there are still difficulties before desirable products come into use. In this regards, present review aims at summarizing recent advances in flexible and transparent TFTs based on ZnO and related materials, and discussing the major challenges in devices fabrication and mechanical strain effects. Finally, we propose several issues to be considered for further investigations.Since there are plenty of reviews on oxide semiconductor TFTs,[11,12,17–21,113,114]to avoid repetition, this review will specifically focus on ZnO and related materials and emphasize the novel device physics and technical problems which only present in flexible and transparent ZnO TFTs.

## 2. Device fabrication

TFTs fabricated on flexible substrates are lightweight, low cost, rugged, flexible, foldable, twistableor even stretchable. However, they are also vulnerable to ambient environment. Therefore, device fabrication and characterizationprocesses may be quite different compared with conventional caseon rigid substrates, such as glass and silicon.[115,116]The most serious problem associated with flexible and transparent (polymer) substrates is their dimension change, which would bring difficulties to sequential alignment. Besides, the mismatch between substrates and films during dimension change would cause strain in the film and thus degrade the materials' quality, or even cracks and delamination which would cause permanent failure. This undesirable dimension change comes from the large difference of coefficient of thermal expansion (CTE), elastic modulus and toughness between the polymer substrates and functional films on them. Other issues with polymer substrates are surface roughness, chemical stabilityand gas permeability, which will be discussed in subsection 2.1.Afterwards, the device physics and technical process in preparing channel, dielectric and electrodes layers will be discussed in subsection 2.2, 2.3 and 2.4, respectively.

### 2.1.Flexible substrates

The properties of polymer substrateswill affect materials quality and carrier transportation behavior and limit maximum fabrication temperature,and thus are of great importance to the flexible transparent devices. As shown in Fig. 3, according to the servicing temperature or glass transition temperature ($T_g$), the polymer substrates can be categorized into three types: conventional polymers ($T_g$< 100 °C), common high temperature polymers (100 ≤$T_g$< 200 °C) and high temperature polymers ($T_g$≥ 200 °C).[117]The most commonly used polymer substrates in the literatures include polyimide (PI), polyarylate (PAR), polyethylene terephthalate (PET), PEN,polyethersulfone (PES), polycarbonate (PC), polyetheretherketone (PEEK), polydimethylsiloxane (PDMS) and so on.

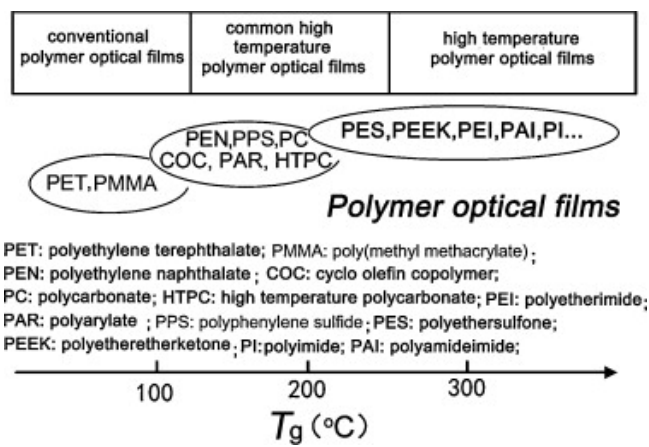

Fig.3 Classification of polymer optical films.According to the servicing temperature, polymer films can be categorized into 3 types: conventional polymers ($Tg$< 100 °C), common high temperature polymers ($100 \leq Tg$< 200 °C) and high temperature polymers ($Tg \geq 200$ °C).[117]

Table 2 lists the basic properties of PI, PEN and PET for each of the three kinds of polymer substrates which are currently widely used.

Table 2 Basic properties of four commonly used flexible transparent substrates.

| Material | $T_g$ (°C) | CTE(ppm °C$^{-1}$) | $\lambda_c$ (nm) | Chemical resistance | Surface roughness |
|----------|-----------|--------------------|------------------|---------------------|-------------------|
| PI | 300 | 12 | 500 | Good | Good |
| PEN | 120 | 20 | 380 | Good | Moderate |
| PET | 80 | 33 | 300 | Good | Moderate |
| PDMS | -120 | 301 | 200 | Good | Poor |

The PI substrate has the highest $T_g$, smallest CTE and surface roughness among all of the flexible substrates. In addition, it also shows good chemical stability in acid, alkali and organic solvents. Although the cut-off wavelength ($\lambda_c$) is in the visible range, which means the PI substrates have deep color and poor optical transmittance as can be seen in Fig. 4, the colorless and optically transparent polyimide (CPI) films are recently developed[117] and used in IGZO TFT fabrication[118]. In the casethatcost is not a problem, the CPI substrates will be the best choice for flexible transparent electronics. The PET substrate has excellent optical transparency and short $\lambda_c$ (Fig. 4), and is currently widely used as the protection layer in various liquid crystal displayers (LCDs), such as televisions, computers and cellphones. The major disadvantage of PET for itsuse in flexible transparent TFTs is its relatively low servicing temperature ($T_g \approx 80$ °C), which may bring challenges in device fabrication, integration and operation. The PEN substrate has higher servicing temperature ($T_g \approx 120$ °C) than PET, but, as a compromise, it appears slightly white color as can be seen in Fig. 4. The PDMS substrate is recently used in flexible TFTs.[27,61,81,92,119] Besides the high optical transmittance and short $\lambda_c$, PMDS also features small elastic modulus (E ≈ 5 MPa), which makes PDMS the perfect substrate for the emerging stretchable electronics.[17,27,61,92,120–122] It is worth noting that ultra-thin (25-100 μm) flexible glass, such as Willow by Corning company,[123] was also regarded as a promising flexible substrate for flexible transparent electronics consideringits higher servicing temperature (~ 500 °C), good resistance to scratch and higher optical transmittance. However, up to now, few researches on flexible glass werereported in the literature.[62,74]

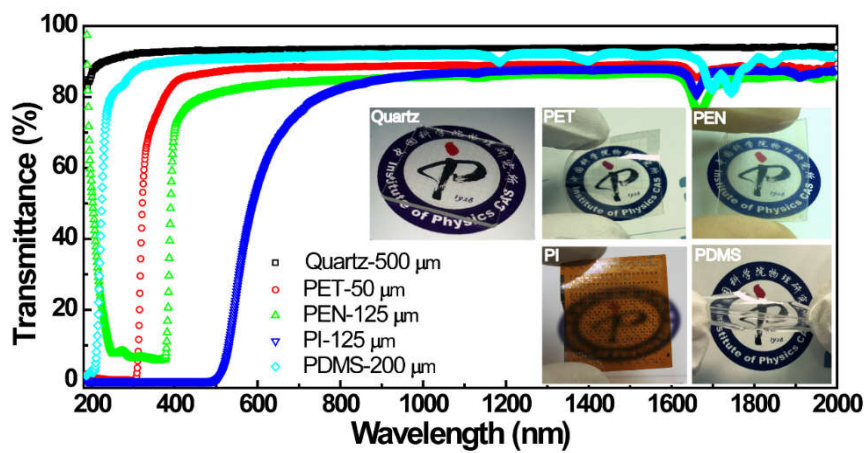

Fig.4 Optical transmittance spectraof four commonly used flexible transparent substrates (PI, PET, PEN& PDMS) in comparisonwith rigid quartz. Insets shows the actual pictures of the substrates.

To decrease the surface roughness and gas permeability, as well as increase the chemical resistance and adhesion with the films, a barrier layer or encapsulation is often involved onto the substrate.[69,83,124] The commonly used encapsulation materials are $Al_2O_3$,[125,126] $Si_3N_4$[58,127] and $SiO_2$[27,82], which are electrical insulating and easy to be grown by chemical vapor deposition with perfect coverage ratio. Encapsulations of stacked layer have also been reported.[67,69,91]

## 2.2. Channel layers

Whatever the substrate is, the device performanceusually depends on the channel layers, especially within 1-2 nm from the interfacial layers.The electron mobility, electron concentration, density of state and interfacial charge would directly influence the field-effect mobility, on/off ratio, sub-threshold swing and turnon voltage. In this part, we summarize the propertiesof ZnO channel layer that only appear in flexible transparent devices.

As described in section 2.1, limited by the utilized flexible substrate, the processing temperature cannot exceed$T_g$, which could be as low as 80 °C.At such a low synthesis temperature, the materials usually appearto be polycrystalline or amorphous. In conventional semiconductors, such assilicon whose conductionband minimum (CBM)and valence band maximum (VBM) are composed of anti-bonding ($sp^3$ σ*) and bonding ($sp^3$ σ) states of Si $sp^3$ orbitals and whose band gap is formed by splitting of the σ*-σ energy levels (Fig. 5a), the carrier transport properties depend critically on the chemical bonding direction (Fig. 5b).This is the reason of the degradation of electron mobility in amorphous silicon ($\mu$<2cm$^2$V$^{-1}$s$^{-1}$) compared with crystalline silicon ($\mu$≈1500cm$^2$V$^{-1}$s$^{-1}$).In contrast to silicon, ZnO has very strong ionicity and electrons transfer from zinc to oxygen atoms. The electronic structure is formed by rising the electronic level in zinc and lowering the electronic level in oxygenthrough the Madelung potential as shown in Fig. 5c. As a result, the CBM in ZnO is primarily formed by the unoccupied spherically symmetric Zn *4s* orbitals and the VBM is primarily formed by the fully occupied axial symmetry O *2p*orbitals.[4,9,128] Thanks to the *s* orbitals formed CBM, the electron transport is not affected significantly by the chemical bond direction and crystal structure randomness.Thatis why high electron mobility occurs in ZnO and related materials in their polycrystalline and even amorphous states, and also why ZnO and related materials are suitable for low temperature process, such as flexible transparent electronics on polymer substrates.

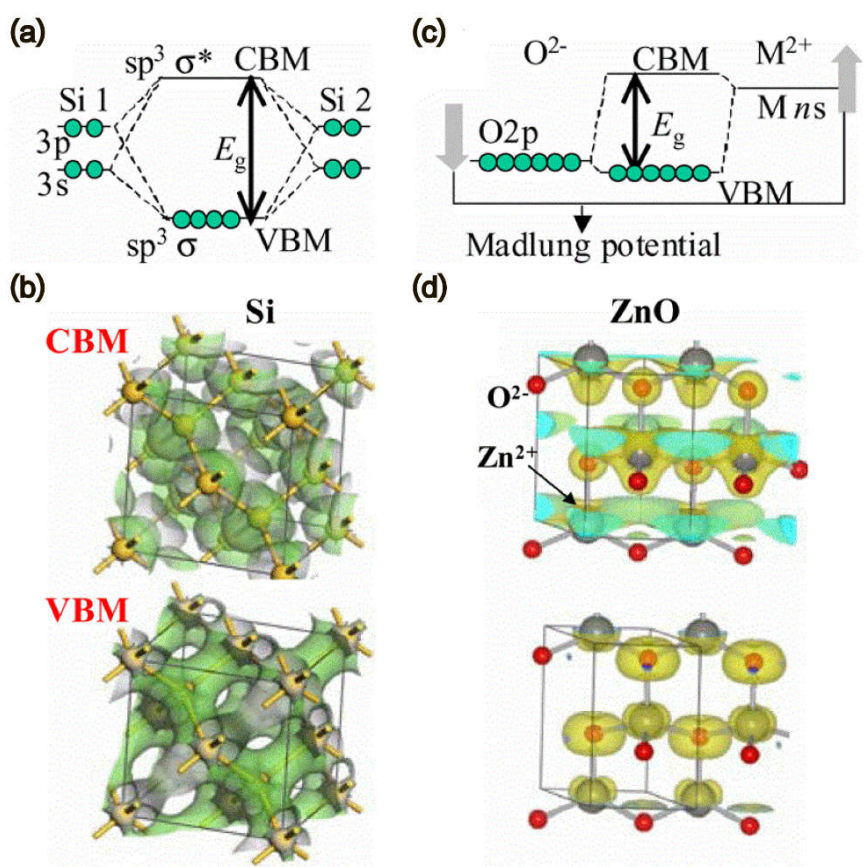

Fig.5 Different formation mechanisms of CBM, VBM and bandgap in silicon (a) and ZnO (c) semiconductors, as well ascrystal structures and CBM/VBM wave functions insilicon (b) and ZnO (d) semiconductors.[128]

To further extend the *s* orbital of ZnO and thus to achieve higher electron mobility, indium (In) and tin (Sn) are often added into ZnO, because $In^{3+}$ and $Sn^{2+}$ have more broad-spreading *5s* orbitals than that of $Zn^{2+}$ *4s*. Ternary alloys such asIZO[52,129] and ZTO[130,131] did show higher mobility than pure ZnO.However, due to the scarcity and toxicity, In is not suitable for the large demands in commercial applications as shown in Fig. 1a-b and the safety use in the healthcare or medical areas as shown in Fig. 1c-d. Sn has been regarded as a more promising candidate to increase the electron mobility in ZnO. At low process temperature, lots of point defectswhich may act as a scattering center and electron trap within the channel layer, are unavoidably induced into ZnO.[132–134]To suppress the concentration of $V_O$, cations,which has a stronger bonding with oxygen,such as gallium (Ga), Sn, aluminum (Al), magnesium (Mg), hafnium (Hf), zirconium (Zr) and Si have been incorporated into ZnO.Quarternary alloys such as InGaZnO[2,4,135], InSnZnO[96,136], MgSnZnO[137],AlSnZnO[138,139], HfInZnO[140,141], ZrInZnO[142] and SiInZnO[143,144] have been reported with improved performance and stability.

As listed in Table 1, the mostcommonly used synthesis technic for ZnO channel layer on polymersremains to be sputtering because its simplicity and effectiveness.Another widely used technic is solution processing techniques, including spin-coating, ink-jet printing, chemical bath deposition and so on.[19,115,95]Although these synthesis technics are important and widely used, there is not much difference ofthe device fabrication techniquebetween rigid and flexible substrates, and they have been described in detail in many reviews[11,12,18,19,21,113,114].Therefore, they will not be discussed here.

On the other hand,roll-to-roll (R2R), sheet-to-sheet (S2S)and roll-to-sheet (R2S) printing technologieshold great promise and offer advantages over classic microfabrication.[61,145]It allows extremely low manufacture cost, large fabrication area, fast printing speed of meters per second[146] and fine feature size of sub-10 μm[147]. The above-mentioned unique advantages make it perfect candidate forapplications in cheap, portable, large-volume and disposable systems, such as radio-frequency identification (RFID) tags, flexible displays and food packaging.[148–150] Fig. 6 shows the scheme of a R2R(Fig. 6a) and R2S(Fig. 6b) gravure printing system, which describes a typical flexible device fabrication procedure. In each of the R2R gravure printing unit, one particular material based ink was selected to print the functional layers, i.e. buffer layers, electrodes, active layers, insulators, passivation, logos and so on.The cells are filled with inks from an ink reservoir and wiped using a doctor blade, after then the inks are transferred from the roll to the flexible substrate.Patterns are defined by recessed cells that are engraved into a roll. After the transfer process, the individual aliquots of inks spread and dry to form the final pattern. After going through every gravure printing unit, multi-functional layers are formed, patterned and aligned to each other as shown in Fig. 6c.

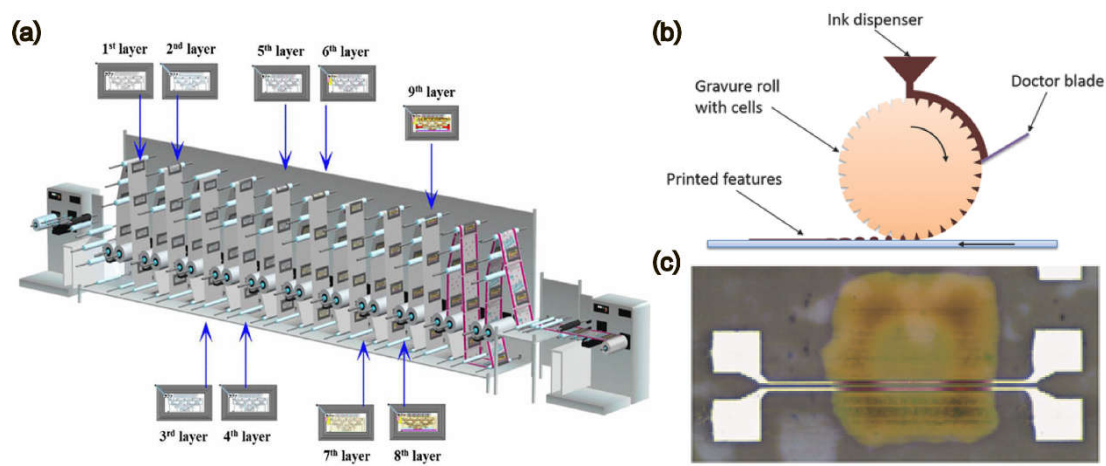

Fig.6 Descriptive scheme for R2R/R2S gravure printing procedure. (a) R2R gravure printing system used to print RF sensor tags and smart packaging on a single line.[145] (b) Overview of a gravure printing process and (c) Optical micrographs of a printed TFT.[147]

## 2.3. Dielectric layers

As the name indicated, transistor (transfer + resistor) is essentially a variable resistor whose resistance is controlled by external electric field, which is generated by gate voltage ($V_g$) within a metal-insulator-semiconductor (MIS) capacitor. Therefore, the quality of the bulk dielectric and interface of semiconductor/dielectrics is of vital importance.Currently, there are three types of dielectric materials that are commonly used in flexible ZnO TFTs in the literatures.

### 2.3.1. Inorganic dielectrics

Silicon dioxide ($SiO_2$) and silicon nitride ($Si_3N_4$) are two kinds of inorganic dielectric materials adopted in a-Si:H and poly-Si TFTs.[24]However, the high deposition temperature above 300 °C for high-quality film by industrialized plasma enhanced chemical vapor deposition (PECVD) hinders their application on flexible substrates.Instead of $SiO_2$[39] and $Si_3N_4$[77],high-$k$ dielectrics are more widely used in flexible transparent ZnO TFTs, because they can be synthesized at low temperature by atomic layer deposition (ALD)[151] or solution processes.We have deposited aluminum oxide ($Al_2O_3$), a typical high-$k$ dielectric material, on rigid silicon, quartz, flexible PEN andPI substrates by ALD at low temperatures of 150 °C, 150 °C,100 °C and 100 °C, respectively.The capacitance–voltage ($C$-$V$) measurements of ZnO/$Al_2O_3$/ITO MIS capacitors at a frequency of 50 KHz on different substrates are shown in Fig. 7.The $Al_2O_3$insulators showed comparable insulating performance on rigid quartz substrateand flexible PEN, PI substrates.The 50-nm-thick $Al_2O_3$ on PEN exhibits a capacitance density of $1.5 \times 10^{-7}$F/cm$^2$ and dielectric constant of 8.47, which is basically equivalent to the dielectric propertiesof the $Al_2O_3$ films fabricated on quartz.

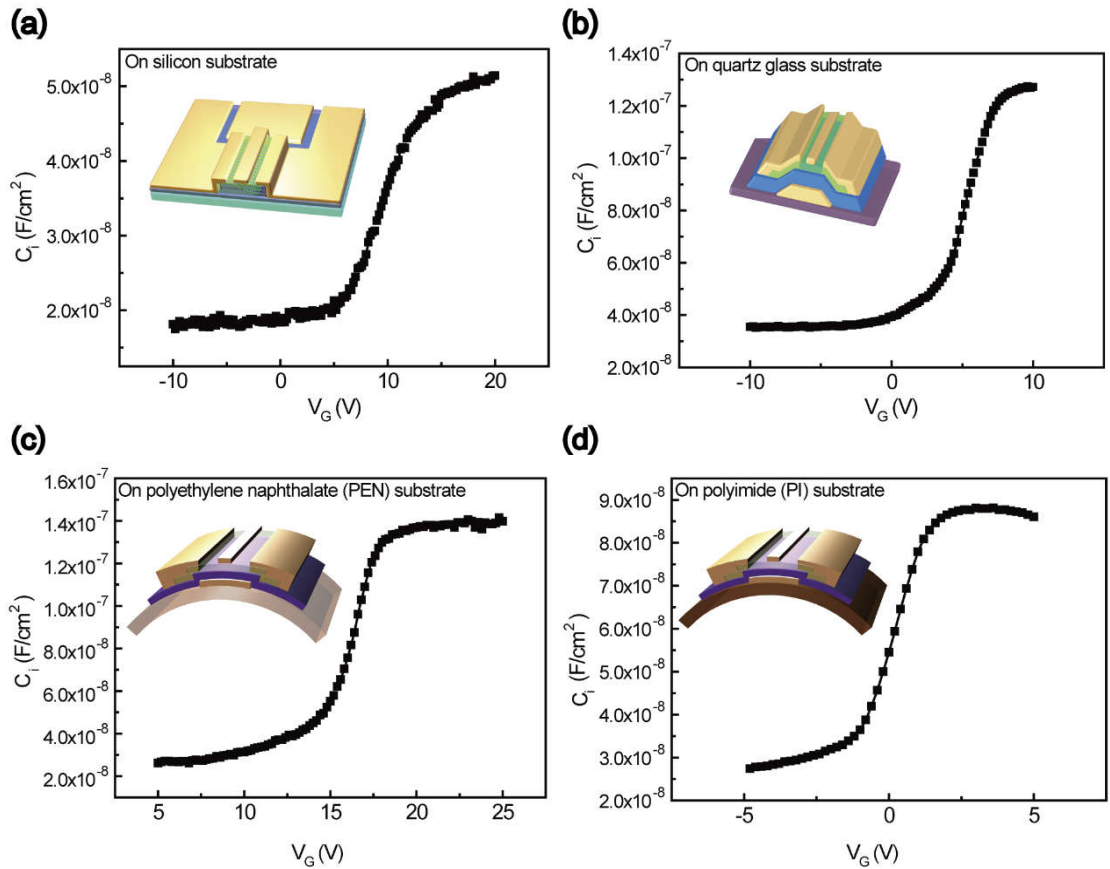

Fig.7 Capacitance-voltage (C-V) measurements of $Al_2O_3$ dielectrics deposited at low temperatures on (a)p$^{++}$-silicon, (b)quartz glass, (c)PEN and (d) PI substrates.

### 2.3.2. Organic dielectrics

Inorganic dielectrics, such as $Al_2O_3$, have exhibit low-temperature fabrication convenience and excellent device performance. However, mechanical failure may occur when the films are under tensile or compressive strains, as shown in Fig. 8. Jen etal. reported that a80-nm thick ALD-grown $Al_2O_3$ film could only sustain a strain level of 0.52%.[152]Xu etal.claimed the critical strain, acritical point at which the film becomes useless, for 200-nm-thick anodized $Al_2O_3$ film on PEN is 0.6%.[66]If the thickness of the flexible substrate is 125 μm, the critical bending radius is approximately 10 mm (See section 3 for more details of mechanical bending), which is not suitable for some flexible, foldable or stretchable applications.

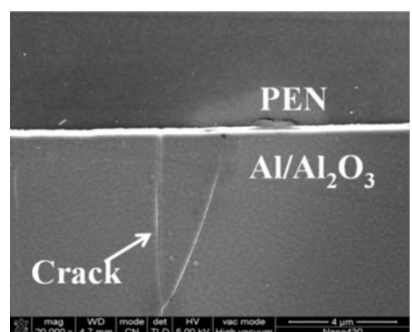

Fig.8 FESEM image of cracks in Al₂O₃ film with a thickness of 200 nm on PEN after 0.6% strain.[66]

Nevertheless, organic dielectric materials,[153] such as poly(4-vinlphenol) (PVP), poly(methyl methacrylate) (PMMA) and polystyrene (PS), can sustain larger strain[91] because the molecules in them are linked through van der Waals bond and/or hydrogen bond which are weak interactions.In addition, polymer dielectrics can be formed by simple and low-cost processes, such as spin-coating and printing. The characteristics of these materials can be tuned by design of the molecular precursors and polymerization reaction conditions, which offer more application opportunities in a wide range of electronic devices. Fig. 9 shows the chemical structures of typical polymeric gate dielectrics.

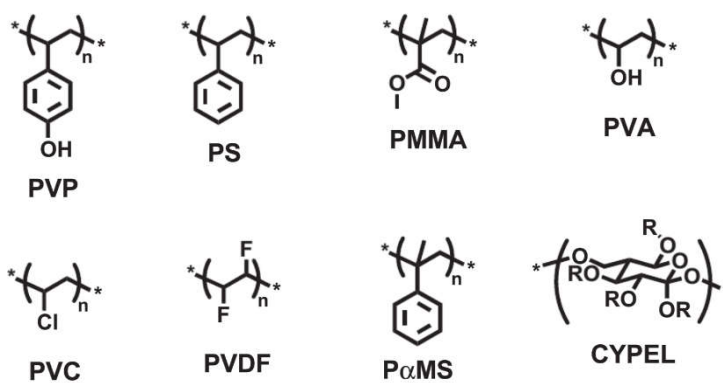

Fig.9 Chemical structures of some typical polymeric dielectrics.[154]

Lai et al. fabricated ultra-flexible IGZO TFT on 125-μm-thick PET substrate with PVP as polymeric gate dielectrics, on which device performance shows no degradation upon bending to strain $\varepsilon$=1.5%.[43]Kim et al. also reported flexible IGZO TFT on PET substrate with PMMA as gate dielectrics.[155] However, they did not perform the bending test evaluation. Up to now, the reports of flexible transparent ZnO TFTs with pristinepolymeric gate dielectrics are still quite limited.

The stable and efficient operation of flexible transparent ZnO TFTs under mechanical stress requires all of the components work stably and reliably. Lai et al. proposed that the polymeric gate dielectrics can also reduce the stress in the IGZO channel layer.[43]The Young's modulus for polymer material is around several GPa. However, it is more than 100 GPa for oxide based inorganic semiconductor. This large difference enables the stress mainly located at the polymer side, leaving the IGZO layer less stressed as shown in Fig. 10.

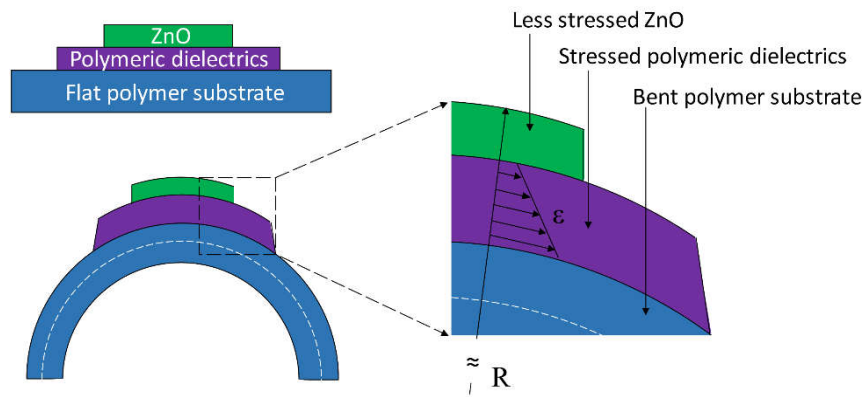

Fig.10 Schematic description of stacked ZnO/polymeric dielectrics/polymer structurein flat and bent states, respectively.In bent state, the stress is mainly located within the polymeric dielectrics layer.

### 2.3.3. Organic/inorganic hybrid dielectrics

Although polymeric dielectrics sustain greater strain than their inorganic counterparts and canrelax the stress in the channel layer,they have some drawbacks.

a. Polymers areusually soft, and thus deposition of channel layer may induce damages inside polymer layer or at the polymer/channel interface, which will significantly influencethe transport behavior of field-modulated electrons.

b. The dielectric constants ($k$) of most polymers are relatively low ($\varepsilon_r$ = 2.5-2.6 for PVP), which would exhibit smaller capacitance at a given thickness than inorganic dielectric materials.

c. The polymeric dielectrics are more hydrophobic than inorganic materials, which is undesirable for directly growth of channel semiconductors.

Besides utilizing stacked organic/inorganic hybrid dielectric gate,[156] these problems could also be solved by introducing inorganic nanoparticles into polymer matrices to form polymer nanocomposites. Lai et al. fabricated a nanocomposite dielectrics by incorporating high-$k$Al$_2$O$_3$ nanoparticles into polymer PVP films as shown in Fig. 11.[68]Al$_2$O$_3$ nanoparticles (size < 50 nm) was added into PVP/poly(melamine-co-formaldehyde) precursorat a concentration from 0.25 to 1.00 wt.%. After stirred overnight, the solution was spin-coated on silver (Ag) gate. The nanocomposite dielectrics was then post-treated with hot plate and ultraviolet illumination. The capacitance of the pristine PVP was 14 nF/cm$^2$ corresponding to a dielectric constant of 3.9. After adding 0.25, 0.5 and 1 wt.%Al$_2$O$_3$ nanoparticles, the capacitance increased to 19, 20 and 22 nF/cm$^2$, the corresponding dielectric constant increasingto 6.1, 7.1 and 8.1.After that, Han et al. introduced lead oxide (PbO), which is a high-$k$ material with $\varepsilon_r$ = 200,[157] into PVP polymer and increased the dielectric constant to 21.2.[158]Along with the improved capacitance, the TFT with nanocomposite dielectrics could sustain a same strainas the device with pristine PVP dielectrics.[43,68]This means the nanocomposite dielectricsinherits the merits from both inorganic high-$k$ material and organic polymer material.In addition, the incorporation of high-*k* nanoparticles into polymeric dielectrics was believed to improve the robustness against the plasma damage during the following sputtering process.[68] In general, organic materials are more hydrophobic than inorganic materials, which is undesirable in the direct growth of inorganic semiconductor materials on the polymeric dielectrics.[159,160] However, few researchers have studied the effect of incorporated nanoparticles on the surface energy of nanocomposite dielectrics.

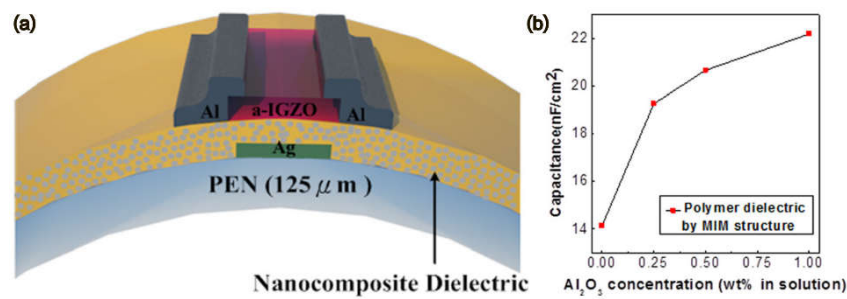

Fig.11Organic/inorganic hybrid nanocomposite dielectrics.(a) Schematic diagram of IGZO TFT with nanocomposite dielectrics. (b) Capacitance of nanocomposite dielectrics in which capacitance increased with $Al_2O_3$ concentration.[68]

## 2.4. Electrode layers

The gate and source/drain electrodes in flexible transparent ZnO TFTs should at least possess three characteristics: low conductive resistivity, high optical transmittance, and good mechanical stability. Yet the most widely used flexible transparent electrodes (FTEs) are transparent conductive oxides (TCOs), represented by ITO, FTO, AZO, GZO, IZO, ZTO and IZTO, as they have wide bandgap, low resistivity and can be deposited at low temperatures.[161,162]However, the mechanical stability of TCO electrodes remains the tough issue to be solved for achieving stable and reliable device operation.[163–166]

Leterrier et al. investigated the effect of ITO thickness on the crack onset strain (COS), the critical strain at failure of the film.[163] The evolution of film mechanical failure under uniaxial strain was recorded as shown in Fig. 12. The as grown ITO film has some microdefects in form of pin-holes (Fig. 12a). Upon a strain of 1.28%, small cracks originated from the microdefects (Fig. 12b). Further increasing the strain to 1.42%, the initial crack propagated from the sites to finite size (Fig. 12c) and the resistivity began to increase as shown in the onset region in Fig. 12e. At higher strain levels, the finite cracks increased and the width spanned to the whole sample (Fig. 12d). As a result, the resistivity increased markedly as can be seen in Fig. 12e. They found that the COS decreased with ITO thickness, which means for a thicker ITO film the safe operating range could be even smaller.

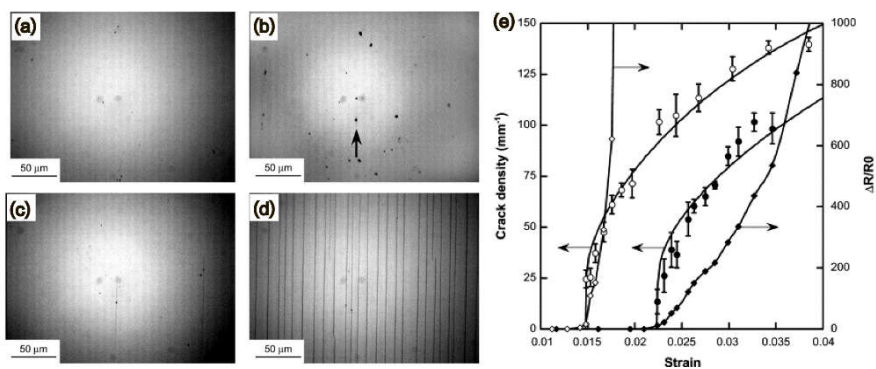

Fig.12 Progressive cracking of a 100 nm thick ITO film on 100 μm thick polyester substrate during tensile loading (along the horizontal direction). Unloaded ITO (a); at 1.28% strain (b), the arrow indicates failure initiation on a coating defect); at 1.42% strain (c); and at 3.42% strain (d). Density of tensile cracks and normalized resistance change during tensile loading of 50 nm (filled symbols) and 100 nm (open symbols) thick ITO (e).[163]

To seek for flexible transparent electrodes which can sustain higher mechanical strains, oxide-metal-oxide (OMO) stacked structure (Fig. 13a)has been proposed. The most commonly used metal material is silver (Ag)and copper (Cu) because oftheir good ductility and high conductivity. What's more, by optimizing the metal layer thickness, the figure of merit (FOM), which can comprehensively characterize the property of transparent electrode, can be improved.[167]By inselting a thin Ag layer into the middle of double 30 nm-thick IZO layers[168–171], the IZO/Ag/IZO stacked electrode showed improved optical transmittance andconductive resistivity (Fig. 13b).[171]Maximum improvement of FOM was obtained with a 12 nm-thick Ag layer (Fig. 13c).More importantly, the mechanical robustness was improved compared with a single ITO layer (Fig. 13d).Other reported stacked electrodes with improved electrical conductivity, optical transmittance and mechanical robustness involve ITO/Ag/ITO[172,173], IZTO/Ag/IZTO[174], GZO/Ag/GZO[175], IZO/Ag/IZO, ZTO/Ag/ZTO[176,177], IZO/Ag/IZO/Ag[170] and ITO/Cu/ITO[173].

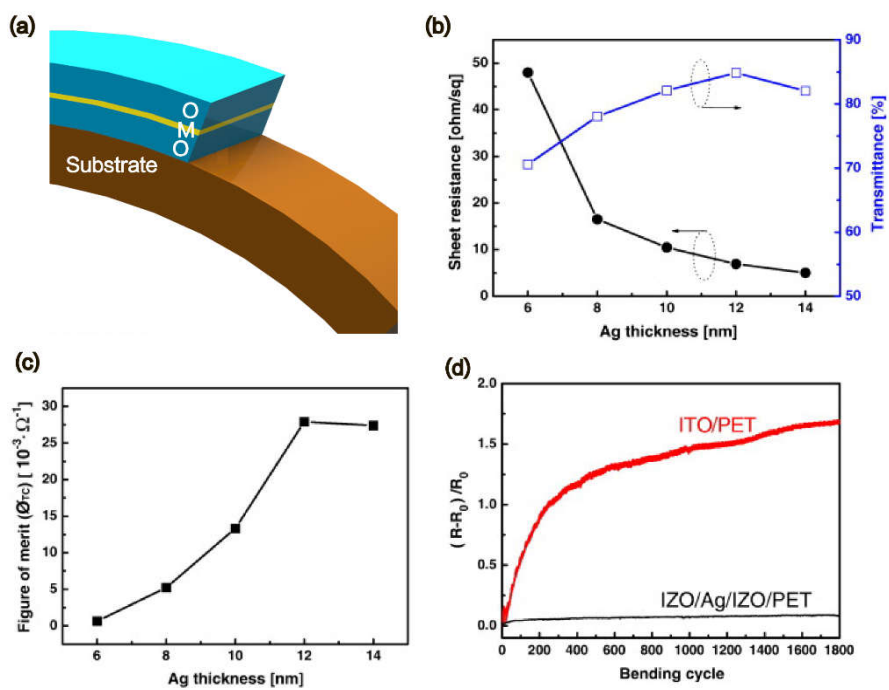

Fig.13OMO stacked electrode.(a) Schematic diagram of a typical OMO stacked electrode. (b)Sheet resistance, transmittance at 550 nm and (c)FOM of an IZO/Ag/IZO multilayer anodes on PET substrates as a function of the Ag thickness. (d) Normalized resistance change after repeated bending as a function of the number of cycles for IZO/Ag/IZO/PET and amorphous ITO/PET sample.[171]

The random meshed Ag nanowire (AgNW)[178–182]electrode is another promising candidate for flexible transparent ZnO TFTs, as Ag forms ohmic contact with ZnO.[104] The unique advantage of Ag nanowire electrodes is their mechanical robustness, because nano-materials can be bent to much smaller radius than conventional "3D" materials.[183]Fig. 14ashows the SEM image of random meshed AgNWs on flexible substrate[182] andFig. 14b shows the resistance change as a function of mechanical strains.[179]Compared with the small COS of ITO shown in Fig. 12, the AgNW flexible transparent electrode processes much better mechanical robustness, with only 3.9 times increase of resistance under 15 % tensile strain and almost no change under 15 % compressive strain.

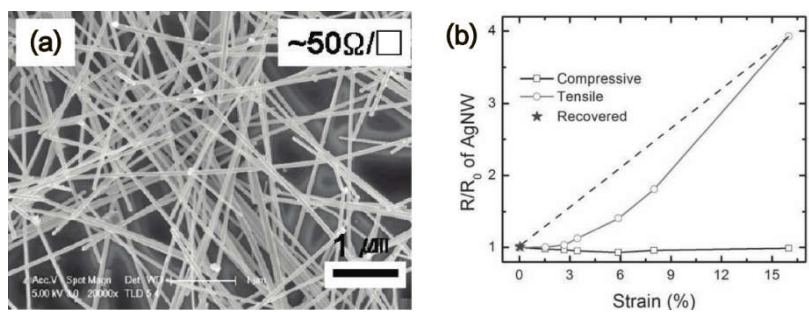

Fig.14Random meshed Ag NW electrode.(a) SEM image of AgNWs on flexible transparent substrate.[182](b) Surface resistance ratio ($R/R_0$) of the AgNW/polymer electrode comparing tensile and compressive strains.[179]

## 3. Bending effect on flexible TFT

Operation of flexible transparent ZnO TFTs often involves mechanical deforming of substrate, so understanding of the evolution of device performance under stress is of fundamental importance to research efforts in this area.[184,185]In fact, the requirement forflexibility of devices is quite different when they are applied in different application areas. In the case of flexible displays, the device may need a large strain tolerance when it was rolled up like a scroll.[6]In skin sensors and wearable electronics, devices may go through small but repeated strains.[22]Despite the differences, the mechanical processes all follow the same basic principles. So in this section, we will review the mechanical fundamentals on flexible transparent ZnO TFTs and focus on the test technique, strain calculation and characterization methods. Finally, the current reports on flexibility test of ZnO TFTs will be briefly depicted.

### 3.1.Bending test systems

The bending test systems reported in the literatures are all laboratory-made and can be roughly classified into two types according to how devices are bent: 1) substrate wrappedaround a rigid rod (Fig. 15a) and 2) arched up under side extrusion (Fig. 15b).In type 1, the probe tips contact well with the electrodes. However, the variation of bending radius is not convenient in this setup. On the contrary, in type 2, the bending radius can be tuned by varying the distance between the two splints. But the probe tips might contact poorly with the electrodes especially when the substrate is thin or soft.[88] Thus, attaching flexible polymer substrate onto a flexible metal sheet might be a good choice to combine good contact and flexibility.

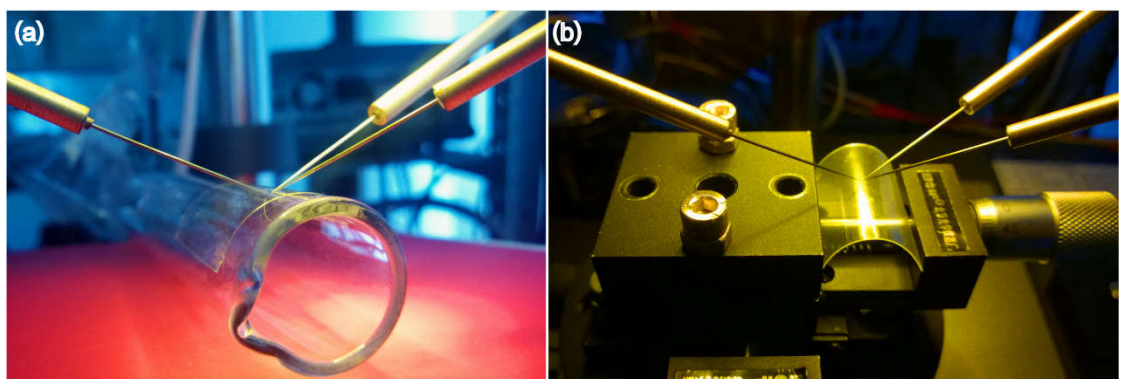

Fig.15 Two types of bending test system. (a) Substrate wrapped around a rigid rod and (b) arched up under side extrusion.

### 3.2.Strain in films

In simplified situation (no Poisson ratio and fabrication-induced strain), the strain ($\varepsilon$) within the films on bending substrates can be roughly obtained through equation 1, under the premise that the substrate is much thicker than the film,orequation 2, with the assumption of neglecting the difference ofthe Young's modulibetween substrate ($Y_s$) and film ($Y_f$).Both the two mechanical models simplified calculation process and suited wellin many situations.[43,68,186]

$$\varepsilon = \frac{d_s}{2R} \tag{1}$$

$$\varepsilon = \frac{d_s + d_f}{2R} \tag{2}$$

where $d_s$ and $d_f$ are the thicknesses of substrate and film, respectively, and $R$ is the bending radius.

For situations which don't meet the premise or assumption, one could go to equation 3[187–189]

$$\varepsilon = \left(\frac{d_s + d_f}{2R}\right) \frac{\left(1 + 2\eta + \chi\eta^2\right)}{\left(1 + \eta\right)\left(1 + \chi\eta\right)} \tag{3}$$

where $\eta = d_f/d_s$ and $\chi = Y_f/Y_s$.

The relationships between $\varepsilon$ and $d_s$, $R$, $\eta$, $\chi$ in equation 1& 3, are visualized in Fig. 16a & 16b (setting $d_s = 100$ μm and $R = 5$ mm). As can be seen in Fig. 16a, at a fixed radius, the strain can be reduced by utilizing a thin substrate. At a fixed $d_s$, the strain increased markedly with small radius, and the critical failure radius increases with $d_s$. Taking $d_f$ and the difference between $Y_f$ and $Y_s$ into account, the strain could be smaller with larger $\eta$ and $\chi$ as can be seen in Fig. 16b.

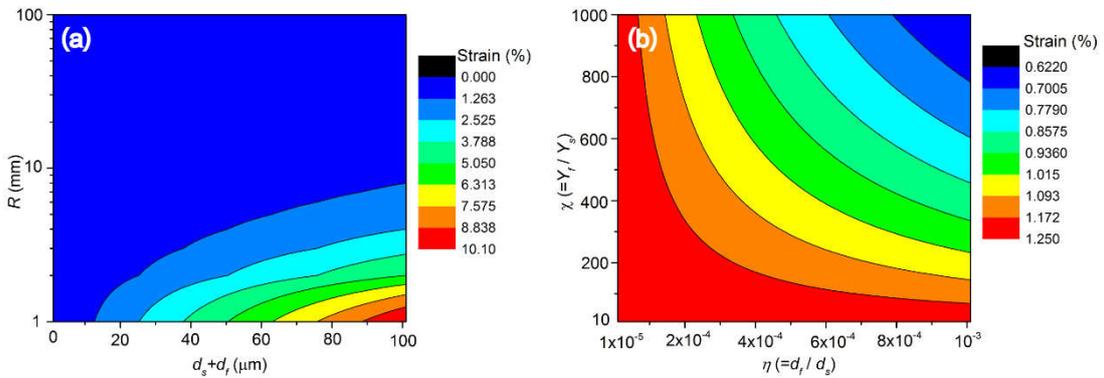

Fig.16 Visualization of relationships between $\varepsilon$ and $d_s$, $R$(a), $\eta$, $\chi$(b).

For a given film (fixed $d_f$ and $Y_f$), besides using thinner and more elastic substrates described above, there are other ways to reduce the strains in functional films. By encapsulating the film and forming an encapsulation/film/substrate sandwiched structure, the strains in film can be further reduced.[189] If the configuration meets

$$Y_s d_s^2 = Y_e d_e^2, \tag{4}$$

where $Y_e$ and $d_e$ are Young's moduli and thickness of the encapsulation layer, respectively, the film without any strain is right in the neutral surface. Consequently, the functional film won't fail to work until the substrate and encapsulation are ineffective. In this approach, Sekitani et al. fabricated ultra-flexible organic TFT with a sandwiched poly-chloro-paraxylylene/pentacene-TFT/PIstructure[190] and Kinkeldei et al. reported sandwiched PI/IGZO TFT/PI structure with a small bending radius of 125 μm.[40] Park et al. proposed that inserting a buffer layer between film and substrate can also reduce the film strains.[191]

### 3.3. Bending directions

In practical applications, the flexible transparent ZnO TFTs might be bent into various forms, and thus the functional films would sustain variouskinds of strains, such as tensile, compressive and twisting.Besides conventional electrical field, the mechanical stress field is also an important issue that must be included in the analysis of flexible electronics. For flexible IGZO TFT, the tensile strain parallel (Fig. 17a) or perpendicular (Fig. 17b) to the current flowhasdifferent effects on the electrical performance.[42]As shown in Fig. 17c, saturation mobility ($\mu_{sat}$) was slightly affected by the strain parallel to the channel up to $\varepsilon$= 0.72%, while $\mu_{sat}$ began to degradedwhen$\varepsilon$> 0.3% under a strain perpendicular to channel. The degradation of both on and off currents of flexible TFT under the perpendicular strain was explained as due to the crack of brittle chromium (Cr) gate. The cracks disconnected parts of gate, leaving some of the channel areas uncontrolled, thus making the off-currentincreased and the on-currentdecreased.Whereas, no crack occurred under the parallel strain when $\varepsilon$< 0.72%. Once $\varepsilon$ exceeds 0.72%, the cracks became unstable and destroyed the device permanently.

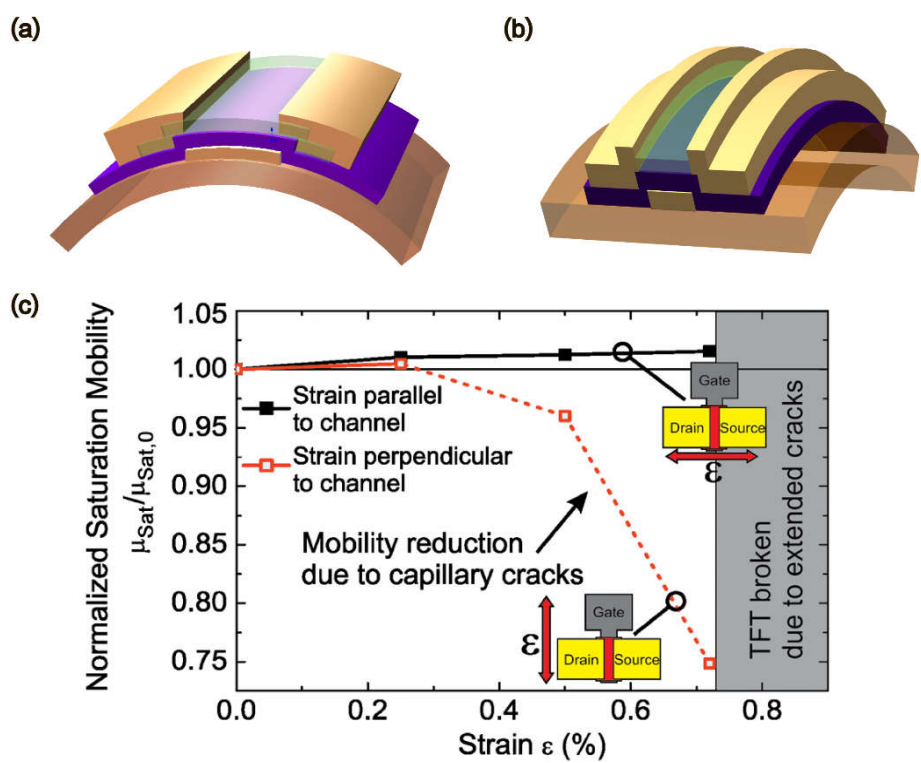

Fig.17Strain directions in reference with the channel direction. Schematic of tensile strain (a) parallel and (b)perpendicular to the channel direction in flexible TFT.(c)Normalized IGZO saturation mobility changes induced by mechanical strains parallel or perpendicular to channel.[42]

Bending the TFTs outwards causesa tensile strain, while bending the TFTs inwards causes acompressive strain. Compared with the tensile strain, the compressive strain has relatively smaller effect on the electrical performance.[192–194]The influence of strain on the electron transport mobility can be described as blow,

$$\frac{\mu_{bent}}{\mu_0} = 1 + m\varepsilon \tag{5}$$

where $\mu_0$ and $\mu_{bent}$are the charge carrier mobility in flat and bent conditions and $m$ is an empirical proportionality constant which depends on the channel material.[195] For IGZO, Münzenriederet al. found that the tensile and compressive strains could induce TFT parameters shifting towards opposite directions as shown in Fig. 18.[72,194]Under a tensile strain, the field mobility and subthreshold swing wereincreased, and the threshold voltage was decreased; while under a compressive strain, the field mobility and subthreshold swing weredecreased, and the threshold voltage was increased. The dependence of electrical performance on the strain types can be explained with the electrical structure change associated with crystal structure deformation.[85,196]

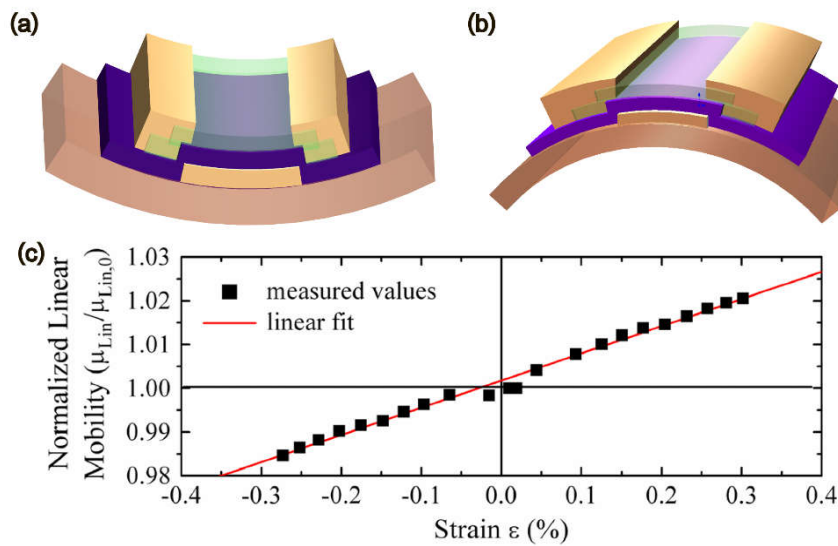

Fig.18Strain types under inward or outward bending. Schematic description of (a) tensile and (b) compressive strains in flexible TFT. (c) Normalized IGZO TFT linear mobility for different tensile andcompressive strains.[194]

### 3.4.Reports on strain test

The maximum operation strain of flexible transparent ZnO TFT depends on the mechanical properties of each of the functional layers.Devices with high robustness that can sustain high intense and repeated bending are always desirable. Table 3 summarized the state of art results on bending test of flexible transparent ZnO TFTs.Most of the polymer substrates listedin Table 3 are PI, PET and PEN with thickness ranging from 1 to several hundred microns. Either wrapped around rigid rods or arched up with two parallel plates, majority of the devices are bent with tensile strain parallel to the current flow (channel length direction). However, compressive strain or strains perpendicular to current flow should also been conducted since they could have different influence on the flexible ZnO TFT performance.[42,194]Generally, smaller bending radii are desirable, because it means more flexibility and serviceability. Bending radius can vary in a large range from 25 μm to 10 cm depending mainly on the thickness of the substrates.Strain is in inverse proportion to bending radius as shown in equation 1, 2 & 3, and is the normalized parameter to describe the mechanical status in functional layers regardless of the substrate thickness. Fatigue test concerns the stability and reliability of flexible devices. Typically, the fatigue tests are of $10^2$ to $10^6$ stressing times.

Table 3 Results on bending test of flexible transparent ZnO TFTs from 2010 to 2016.

| Substrate | Setup | Paral. or perp. to current flow | Compressive or tension | Bending degree | | Bending times | Ref. | Year |
|---|---|---|---|---|---|---|---|---|
| | | | | Radius(mm) | Strain(%) | | | |
| 50 μm PI | Arch | -- | Tensile | 10/5 | 0.56/1.06 | -- | Cherenack et al.[26] | 2010 |
| 220 μm PET | -- | Paral. | -- | 40 | 0.28 | -- | Liu et al.[35] | 2010 |
| 50 μm PI | Arch | -- | Tensile | 0.25 | 6.35 | 100 | Song et al.[29] | 2010 |
| 50 μm PI | Arch | Paral. | Compressive | 0.125 | -- | -- | Kinkeldei et al.[40] | 2011 |
| 50 μm PI | Wrap | Both | Tensile | 3.5 | 0.72 | 20000 | Münzenrieder et al.[42] | 2012 |
| 125 μm PEN | Wrap | -- | -- | 4 | 1.5 | 100 | Lai et al.[43] | 2012 |
| 125 μm PI | -- | -- | Tensile | 5 | 1.25 | -- | Kim et al.[45] | 2012 |
| 50 μm PI | Wrap | Paral. | Tensile | 7 | 0.36 | -- | Yang et al.[51] | 2013 |
| 50 μm PI | Wrap | Paral. | Tensile | 3.5 | 0.72 | -- | Münzenrieder et al.[58] | 2013 |
| 50 μm PI | Wrap | Paral. | Tensile | 3.5 | 0.72 | -- | Münzenrieder et al.[59] | 2013 |
| 50 μm PI | Wrap | Paral. | Tensile | 5 | 0.5 | -- | Zysset et al.[60] | 2013 |
| PET | Arch | -- | Both | 4 | -- | 6 | Ji et al.[47] | 2013 |
| PET | -- | -- | -- | 20 | -- | -- | Zhou et al.[52] | 2013 |
| 50 μm PI | -- | -- | Tensile | 5 | 1 | 10000 | Hong et al.[49] | 2013 |
| 100μm glass | -- | Perp. | -- | 70 | -- | 100 | Dai et al.[62] | 2013 |
| 200μm PET | Arch | -- | Both | -4.2/4.3 | -- | 10000 | Kim et al.[48] | 2013 |
| 50 μm PEN | Arch | -- | Both | 4 | 0.6 | $10^6$ | Xu et al.[66] | 2014 |
| 120 μm PI | -- | -- | -- | 10 | -- | -- | Wee et al.[64] | 2014 |
| 125 μm PEN | Wrap | perp. | Tensile | 4 | 1.56 | 100 | Lai et al.[68] | 2014 |
| PES | -- | -- | Both | 18 | 0.55 | -- | Park et al.[63] | 2014 |
| 1 μm parylene | Wrap | Paral. | Tensile | 0.05 | 0.4 | -- | Salvatore et al.[23] | 2014 |
| 70 μm glass | Wrap | Perp. | Tensile | 40 | 0.09 | -- | Lee et al.[74] | 2014 |
| 50 μm PI | Wrap | Paral. | Tensile | 4 | 0.4 | -- | Münzenrieder et al.[73] | 2014 |
| 1 μm PI | Wrap | Both | Compressive | 0.025 | -- | -- | Karnaushenko et al.[16] | 2015 |
| PC | -- | -- | -- | 20 | -- | 100 | Hsu et al.[86] | 2015 |
| 10 μm PI | -- | -- | Tensile | 2.6 | -- | 1000 | Honda et al.[78] | 2015 |
| 5 μm PI | Wrap | Paral. | Tensile | 3.5 | 0.07 | 50000 | Li et al.[75] | 2015 |
| PET | Wrap | Paral. | Tensile | 10 | -- | 1000 | Liu et al.[76] | 2015 |
| 25 μm PEN | -- | Both | Tensile | 2 | 0.75 | 4000 | Tripathiet al.[85] | 2015 |
| 125 μm PEN | Arch | Paral. | Tensile | 3.3 | 1.9 | 10000 | Park et al.[83] | 2015 |
| 50 μm PI | Wrap | perp. | Tensile | 5 | 0.48 | -- | Petti et al.[84] | 2015 |
| 17 μm PI | Arch | Both | Tensile | 1.5 | 3.5 | 10000 | Park et al.[91] | 2016 |
| 3 μm PI | Wrap | -- | Tensile | 0.15 | -- | -- | Kim et al.[93] | 2016 |
| 200 μm PES | Both | Paral. | Tensile | 12 | 0.8 | 2000 | Oh et al.[94] | 2016 |
| 500 μm PDMS | Wrap | -- | Tensile | 15 | -- | -- | Jung et al.[92] | 2016 |
| 20 μm PI | Wrap | -- | Tensile | 100 | -- | -- | Zhang et al.[89] | 2016 |
| 125 μm PEN | Arch | Paral. | Tensile | 8 | 0.78 | -- | Zhang et al.[88] | 2016 |

## 4. Conclusion and perspectives

The flexible and transparent electronics have drawn particular attentions in electronic material, device and circuit areas, especially in the last thirteen

years.Mechanical deformation capability, light weight, low cost and other unique advantages make it rising above conventional electronic circuits which are based on rigid substrates, like silicon and glass.The inorganic flexible transparent electronic devices and products are mainly based on ZnO and related materials, owing to their high electrical properties, good optical transparency and low-synthesis temperature. On the contrast, the requirement of uniform material growth in large area at low-temperature rules out other inorganic semiconductors, such as Si, GaN and SiC, which need elevated temperatures for good materials quality. This paper gives a brief introduction on recent advances in flexible transparent TFTs based on ZnO and related materials, and discusses about several important issues of device physics and fabrication technology concerning about the substrate, electrodes, channel and dielectric layer in section 2. The operation and evaluation of strain test are highlighted in section 3.

The emergency of flexible transparent electronics has also boosted the development of solution process technics, especially roll-to-roll printing and other printing technics.[19,21,115,162,181]The inks used in printing technics would need to dissolve the zinc precursors into solvents,such as dissolving zinc hydroxide ($Zn(OH)_2$) in aqueous ammonia ($NH_4OH$)[37,49], zinc acetate dihydrate [$Zn(CH_3COO)_2$]in 2-methoxyethanol ($CH_3OCH_2CH_2OH$)[28,46,93,95,197–200] and zinc chloride ($ZnCl_2$) in ethylene glycol ($C_2H_6O_2$)[62].Thus, a high-temperature (> 300 °C) post annealing is essentialto fully decompose the organic components and produce a pure-phase metal oxide from the solution phase, which is not compatible with most of the flexible substrates.[49]To solve this problem, some low-temperature post treatment is conducted on solution processed ZnO TFTs, including microwave annealing[37,197–199,201] and ultraviolet photo annealing[46,202,203]. However, most of the researches are based on rigid silicon or glass substrates, very limited reports on flexible substrates.[63] So, we suggest that low-temperature post-treatmentcould be one promising direction for solution processing, especially printing, technics.The flexibility also involved mechanical stability issue, which never presents itself in traditional rigid devices. The strain in the film would induce mechanical stress field which could change the crystal, and thus electrical, structure and even crack the oxide films.[204] Critical parameters such as energy band, mobility, dielectricity, and thermal conductivity might also be affected. Thus, coupling of multi-physics involving mechanical stress field needs to be rigorously studied. Finally, the commercialized application still calls for the long-term stable, repeatable, reliable, large-area uniform, and yet low lost, techniques.

Despite the to-be-solved issues, given the rapid development speed at present, we have plenty of reasons to keep optimistic for the coming era of flexible transparent technology.